\documentclass[twocolumn,preprintnumbers, amssymb,amsmath,aps,floatfix,prd,nofootinbib,superscriptaddress,showpacs]{revtex4-1}

\usepackage{epsfig}
\usepackage{bm}
\usepackage{amssymb}
\usepackage{amsmath}
\usepackage{color}
\usepackage{subfigure}
\usepackage[colorlinks,
            linkcolor=blue,
            anchorcolor=black,
            citecolor=blue
            ]{hyperref}

\begin{document}
\title{Anisotropy in Dijet Production in Exclusive and Inclusive Processes }

\author{Yoshitaka Hatta}
\affiliation{Physics Department, Building 510A, Brookhaven National Laboratory, Upton, NY 11973, USA}
\affiliation{RIKEN BNL Research Center,  Brookhaven National Laboratory, Upton, New York 11973, USA}

\author{Bo-Wen Xiao}
\affiliation{School of Science and Engineering, The Chinese University of Hong Kong, Shenzhen 518172, China}

\author{Feng Yuan}
\affiliation{Nuclear Science Division, Lawrence Berkeley National
Laboratory, Berkeley, CA 94720, USA}

\author{Jian Zhou}
\affiliation{\normalsize\it  Key Laboratory of Particle Physics and
Particle Irradiation (MOE),Institute of Frontier and
Interdisciplinary Science, Shandong University (QingDao), Shandong
266237, China }


\begin{abstract}
We investigate the effect of soft gluon radiations on the azimuthal angle correlation between the total and relative momenta of two jets in inclusive and exclusive dijet processes. We show that the final state effect induces a sizable $\cos(2\phi)$  anisotropy due to gluon emissions near the jet cones. The phenomenological consequences of this observation are discussed for various collider experiments, including diffractive processes in ultraperipheral $pA$ and $AA$ collisions, inclusive and diffractive dijet production at the EIC, and inclusive dijet in $pp$ and $AA$ collisions at the LHC.
\end{abstract}
\maketitle


{\it 1. Introduction.} Dijet productions are the most abundant events in hadronic collisions and have been under intensive investigations from both experiment and theory sides. Typically the two final state jets are produced in the so-called correlation configuration, namely, back-to-back in the transverse plane with nearly balanced transverse momenta~\cite{Abazov:2004hm,Abelev:2007ii,Khachatryan:2011zj,daCosta:2011ni,Aad:2010bu,Chatrchyan:2011sx,Adamczyk:2013jei,Aaboud:2019oop}. Deviations from the perfect back-to-back configuration inform important strong interaction physics, and, in particular, the non-perturbative structure of the nucleon and nucleus~\cite{Boer:2003tx,Marquet:2007vb,Dominguez:2010xd,Dominguez:2011wm,Boer:2010zf,Metz:2011wb,Dumitru:2015gaa,Boer:2017xpy,Boer:2016fqd,Xing:2020hwh,Altinoluk:2015dpi,Hatta:2016dxp,Mantysaari:2019csc,Mantysaari:2019hkq}. It is expected to reveal the medium property by studying de-correlation of dijet in heavy ion collisions as well~\cite{Mueller:2016gko,Mueller:2016xoc,Chen:2016vem,Chen:2016cof,Chen:2018fqu,Tannenbaum:2017afg,Adamczyk:2017yhe,Gyulassy:2018qhr}. 

\begin{figure}[tbp]
\begin{center}
\includegraphics[width=3.5cm]{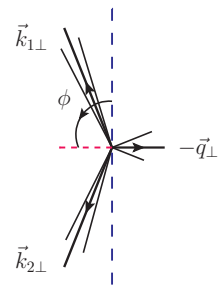}
\end{center}
\caption[*]{Dijet in transverse plane perpendicular to the beam direction at hadron colliders. Their total transverse momentum $\vec{q}_\perp=\vec{k}_{1\perp}+\vec{k}_{2\perp}$ is much smaller than the individual jet momentum $\vec{P}_\perp=(\vec{k}_{1\perp}-\vec{k}_{2\perp})/2$. Angular distribution between $q_\perp$ and $P_\perp$ has an anisotropy due to soft gluon radiation associated with the final state jet with a non-zero $\langle \cos(2\phi)\rangle$. }
\label{fig:dijet0}
\end{figure}

Let $\vec{q}_\perp=\vec{k}_{1\perp}+\vec{k}_{2\perp}$ be the total transverse momentum of the two jets. We also define $\vec{P}_\perp=(\vec{k}_{1\perp}-\vec{k}_{2\perp})/2$ as the leading jet transverse momentum, see Fig.~\ref{fig:dijet0}. The correlation limit is defined by the condition $q_\perp\ll P_\perp\sim k_{1\perp}\sim k_{2\perp}$. In general, the azimuthal angular correlation between $\vec{q}_\perp$ and $\vec{P}_\perp$ is isotropic. However, anisotropic distribution can be generated from nontrivial correlations in the transverse momentum dependent parton distribution (TMD) associated with the incoming hadrons. This provides a unique probe to the novel tomography imaging of the nucleon and nucleus in dijet production.

Among the proposed observables, of particular interest is the $\cos(2\phi)$ anisotropic correlation in exclusive diffractive dijet production \cite{Hatta:2016dxp} where $\phi$ is the angle between $\vec{P}_\perp$ and the target recoil momentum $\vec{\Delta}_\perp$. Such a correlation provides a unique access to the so-called elliptic gluon Wigner distribution of the target~\cite{Hatta:2016dxp,Hagiwara:2016kam,Zhou:2016rnt}. Partly motivated by this observation (see, also, \cite{Hagiwara:2017fye}), very recently, the CMS collaboration has reported a preliminary measurement of this anisotropy in ultraperipheral collisions (UPC) at the LHC~\cite{CMS:2020ekd}. Using $q_\perp$ as a proxy for $\Delta_\perp$, CMS reported a significant $\cos(2\phi)$ asymmetry. Intrigued by this result, in this paper, we will perform a detailed analysis of this anisotropic distribution, focusing mainly on the soft gluon radiation contributions. 

Soft gluon radiation contributions to the azimuthal angular anisotropy of $\cos(2\phi)$ and their resummation for various processes have been investigated in Refs.~\cite{Boer:2006eq,Berger:2007si,Bacchetta:2008xw,Bacchetta:2019qkv,Nadolsky:2007ba,Catani:2010pd,Catani:2014qha,Catani:2017tuc}. For the jet-related processes, they were first studied in Refs.~\cite{Catani:2014qha,Catani:2017tuc} and it was found that their contributions are not power suppressed and can be resummed to all orders in perturbation theory. In this paper, we will follow these developments and apply to dijet production processes. For azimuthally symmetric distributions, the soft gluon resummation has been derived in Refs.~\cite{Banfi:2003jj,Banfi:2008qs,Hautmann:2008vd,Mueller:2013wwa,Sun:2014gfa,Sun:2015doa,Hatta:2019ixj,Liu:2018trl,Liu:2020dct}. 

The physical picture is as follows. Soft gluons emitted from the final state jets tend to be aligned with the jet directions. Since gluons emitted too close to the jet axis become part of the jet, what matters is the emissions slightly outside the jet cones, and $\vec{q}_\perp$ is essentially the recoil momentum against these gluons. This means that $\vec{q}_\perp$ also tends to point to jet directions on average, resulting in a positive $\langle \cos (2\phi)\rangle$. When $q_\perp\ll P_\perp$, one needs to perform the resummation of logarithms $\alpha_s^n(\ln P_\perp/q_\perp)^m$ which can be done in the TMD framework. 
We shall demonstrate that this simple picture can explain, at least partly, the observation by the CMS collaboration. 

Our result in this paper will have a broad impact on the tomographic study of nucleons and nuclei at the future electron-ion collider (EIC)~\cite{Accardi:2012qut,Proceedings:2020eah}. Various anisotropy observables have been proposed to study the novel structure of nucleon and nucleus~\cite{Boer:2010zf,Metz:2011wb,Dumitru:2015gaa,Boer:2017xpy,Boer:2016fqd,Xing:2020hwh,Altinoluk:2015dpi,Hatta:2016dxp,Mantysaari:2019csc,Mantysaari:2019hkq}. The results in these previous works have to be re-examined.

{\it 2. Diffractive Photoproduction of Dijet.}
We first study the diffractive photoproduction of dijets, $\gamma A\to q\bar q+A$. The photon fluctuates into a quark-antiquark pair which then scatters off the nucleon and nucleus target and forms a final state dijet with momentum $k_1$ and $k_2$. This process can be studied, for example, in ultraperipheral $pA$ and $AA$ collisions and the planned EIC in the future.

Let us define the soft factor $S_J(q_\perp)$ as the probability for the dijet to emit total transverse momentum $q_\perp$ outside the jet cones. Implicitly, it should also depend on the dijet relative transverse momentum $P_\perp=\frac{1}{2}(k_{1\perp}-k_{2\perp})$. The dijet cross section then reads
\begin{equation}
\frac{d\sigma}{d^2P_\perp d^2q_\perp}=\int d^2q'_\perp \frac{d\sigma_0}{d^2P_\perp d^2q'_\perp} S_J(q_\perp-q'_\perp),
\end{equation}
where $\sigma_0$ is the cross section without soft radiations. 
We assume that the two jets are back-to-back in azimuth, and consider the soft regime $|q_\perp| \ll |P_\perp|$. To lowest nontrivial order, $S_J$  is given by the standard eikonal formula
\begin{equation}
S_J(q_\perp)=\delta(q_\perp)+\frac{\alpha_s}{2\pi^2}\int dy_g \left(\frac{k_1\cdot k_2}{k_1\cdot k_g k_2\cdot k_g}\right)_{\vec{q}_\perp=-\vec{k}_{g\perp}},\label{eq:sgk1k2}
\end{equation}
where $k_g$ is the soft gluon momentum emitted from the dijet and $y_g$ is its rapidity. A power-counting argument shows that $S_J(q_\perp)\sim 1/q_\perp^2$ at small $q_\perp$, and we neglect power corrections  $(q_\perp/P_\perp)^n$ to this leading behavior such as coming from gluon emissions from the $t$-channel Pomeron (see, e.g., \cite{Boussarie:2014lxa}). We shall be particularly interested in the $\phi$-distribution where $\phi$ is the azimuthal angle between $q_\perp$ and $P_\perp$. In general one can expand
\begin{eqnarray}
S_J(q_\perp)= S_{J0}(|q_\perp|)+2\cos(2\phi)S_{J2}(|q_\perp|)+\cdots\ ,
\end{eqnarray}
where for simplicity higher harmonics are neglected in this work.  To ${\cal O}(\alpha_s)$, $S_{J0,J2}$ can be obtained by calculating the Fourier coefficients 
\begin{equation} 
\int_0^{2\pi} d\phi \int_{-\infty}^\infty dy_g \frac{k_1\cdot k_2}{k_1\cdot k_g k_2\cdot k_g}\{1, \cos (2\phi)\}.
\end{equation}
There are collinear divergences when $k_g$ is parallel to either $k_1$ or $k_2$. They are factorized into the jet functions associated with the two final state jets. In practice, we introduce the theta-function constraints
\begin{eqnarray}
 \Theta(2\cosh(y_1-y_g)-2\cos \phi-R^2) \nonumber \\
 \approx 1-\Theta(R^2-(y_1-y_g)^2-\phi^2),
 \end{eqnarray}
where $R$ is the jet radius, and similarly for the second jet. In the narrow jet approximation $R\ll 1$~\cite{Aversa:1988vb,Mukherjee:2012uz}, we obtain
\begin{equation}
    S_{J0}(q_\perp)=\delta(q_\perp)+\frac{\alpha_0}{\pi}\frac{1}{q_\perp^2} \ ,~~~
    S_{J2}(q_\perp) =\frac{\alpha_2}{\pi}\frac{1}{q_\perp^2} \ ,
\end{equation}
where 
\begin{eqnarray}
\alpha_0=\frac{\alpha_sC_F}{2\pi}2\ln\frac{a_0}{R^2}\ ,~~~
\alpha_2=\frac{\alpha_sC_F}{2\pi}2\ln\frac{a_2}{R^2}\ . \label{eq:lo}
\end{eqnarray}
$a_0$ depends on the rapidity difference $\Delta y_{12}=|y_1-y_2|$ as $a_0=2+2\cosh(\Delta y_{12})$. A closed analytic expression is not available for $a_2$ except in the two limiting cases $\Delta y_{12}=0,\infty$, but it has very mild dependence on $\Delta y_{12}$. It increases slightly from $a_{2}=1/4$ for $\Delta y_{12}=0$ to $a_{2}=1/e$ for $\Delta y_{12}\to \infty$. 

We observe that the $\cos(2\phi)$ term is not power suppressed, which is consistent with the finding in Ref.~\cite{Catani:2017tuc}. In the small-$R$ limit, the ratio $\alpha_2/\alpha_0$ is essentially unity, and the $\phi$ distribution reduces to  two delta functions around $0$ and $\pi$. However, for realistic values of $R$, typically $\alpha_2 /\alpha_0\lesssim 0.2$ because $a_0\gg a_2$.  

To have a reliable prediction, we need to perform the resummation of higher order contributions  to $S_J$ \cite{Hatta:2019ixj}. The constraint $q_\perp = -k_{1g\perp}-k_{2g\perp}-\cdots$ from multiple gluon emissions in momentum space can be conveniently deconvoluted by Fourier-transforming to $b_\perp$-space,
\begin{eqnarray}
    \widetilde{S}_{J}(b_\perp)&=&\int d^2q_\perp e^{i\vec{q}_\perp\cdot \vec{b}_\perp} S_J(q_\perp) \nonumber\\
    &=& \widetilde{S}_{J0}(|b_\perp|)-2\cos(2\phi_b)\widetilde{S}_{J2}(|b_\perp|)+\cdots  \ ,
\end{eqnarray}
where $\phi_b$ represents the angle between $\vec{b}_\perp$ and $\vec{P}_\perp$. Again we have only kept the leading two harmonics. To one-loop, we find
\begin{eqnarray}
\widetilde{S}_{J0}(b_\perp)=1+\alpha_0\ln(\mu_b^2/P_\perp^2)\ ,~~
\widetilde{S}_{J2}(b_\perp)=\alpha_2 \ , \label{find}
\end{eqnarray}
where $\mu_b=2e^{-\gamma_E}/b_\perp$ with $\gamma_E$ being the Euler constant. In the above equation, we have taken into account the virtual diagram contribution to $S_{J0}$ whose natural scale is the jet momentum $P_\perp$. Note that in the $S_{J2}$ term, the Fourier transformation does not introduce IR divergence. We extend (\ref{find}) to all orders by exponentiating to the Sudakov form factor,
\begin{eqnarray}
\widetilde{S}_{J0}(b_\perp)=e^{-\Gamma_0(b_\perp)} \ ,~~
\widetilde{S}_{J2}(b_\perp)=\alpha_2e^{-\Gamma_0(b_\perp)} \ , \label{sud}
\end{eqnarray}
where $\Gamma_0(b_\perp)=\int_{\mu_b^2}^{P_\perp^2}\frac{d\mu^2}{\mu^2} \alpha_0$. Here and in what follows, we neglect the non-global logarithms (NGLs) \cite{Dasgupta:2002bw,Banfi:2008qs} which are not significant in the kinematics of our interest. 

\begin{figure}[tbp]
\begin{center}
\includegraphics[width=6.0cm]{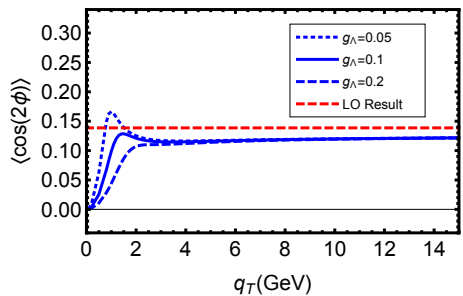}
\end{center}
\caption[*]{Anisotropy in diffractive dijet production  $\gamma+A\to q\bar q+A$ in ultra peripheral heavy ion collisions at the LHC. The kinematics correspond to the CMS measurements~\cite{CMS:2020ekd} with the leading jet $P_\perp=35\rm GeV$, the two jets are at the same rapidity $\Delta y_{12}=0$ and the jet size $R=0.4$. The figure shows $\langle \cos(2\phi)\rangle$ as a function of $q_\perp$, where $\phi$ is the azimuthal angle between $q_\perp$ and $P_\perp$. The leading order result (LO) represents the ratio $\alpha_2/\alpha_0$ of Eq.~(\ref{eq:lo}), whereas full resummation depends on a non-perturbative parameter $g_\Lambda$ (see the main text for the description).}
\label{fig:diffv2qt}
\end{figure}

In the fixed coupling case, the Fourier transform of (\ref{sud}) can be obtained analytically, and we find that $\langle \cos(2\phi)\rangle=S_{J2}(q_\perp)/S_{J0}(q_\perp)$ remains a constant independent of $q_\perp$, but its value is reduced by a factor of $(1-\alpha_0)$ compared to the leading order result $\alpha_2/\alpha_0$. In the running coupling case, we proceed numerically and adopt the $b_*$-prescription~\cite{Collins:1984kg}: $\Gamma_0(b_\perp)\Longrightarrow \Gamma_0(b_*)+g_\Lambda b_\perp^2$, where $b_*=b_\perp/\sqrt{1+b_\perp^2/b_{\rm max}^2}$ with $b_{\rm max}=1.5$~GeV$^{-1}$. The parameter $g_\Lambda$ is applied to include  non-perturbative effects at large $b_\perp$, which should be order $\Lambda_{\rm QCD}^2$. In the numerical  calculations, we take three different values $g_\Lambda=0.05$, $0.1$, $0.2~\rm GeV^2$. In Fig.~\ref{fig:diffv2qt}, we show our prediction for $\langle \cos(2\phi)\rangle$ for diffractive dijet production $\gamma A\to q\bar q+A$, with $P_\perp=35\, \rm GeV$, $R=0.4$ and $\Delta y_{12}=0$. For this jet size, the narrow jet approximation used in the above derivations is reasonably good~\cite{Aversa:1988vb,Mukherjee:2012uz}. We have explicitly verified this by comparing  Eq.~(\ref{eq:lo}) with the exact numerical results. In the figure, we have also plotted the leading order result $\alpha_2/\alpha_0\approx 0.14$ for comparison. We see that the anisotropy does not change dramatically with $q_\perp$ in most  kinematics except in the very small $q_\perp$ region, where the prediction is sensitive to the non-perturbative contributions as expected.   

In the above result, we have neglected the possible contribution from the elliptic gluon Wigner distribution~\cite{Hatta:2016dxp}. The anisotropy $\langle \cos (2\phi)\rangle$ due to the elliptic Wigner is at most a few percent \cite{Hagiwara:2016kam,Mantysaari:2019csc}, which is further suppressed after convoluted with the  symmetric part $S_{J0}$  \cite{Hatta:2019ixj}. It is thus negligible compared to the result in  Fig.~\ref{fig:diffv2qt}. 

In Ref.~\cite{Mantysaari:2019hkq}, a significant $\cos(2\phi)$ anisotropy was also found from a leading order color-glass-condensate calculation in the kinematics of $q_\perp\gg P_\perp$, which is opposite to the correlation limit considered in this paper. Comparison of these $\cos(2\phi)$ anisotropies will help us to understand the underlying physics in dijet processes.

{\it 3. Inclusive Dijet in $\gamma p$ Collisions.}
Next we turn to inclusive dijet photoproduction $\gamma p \to jjX$. We shall  show that, in contrast to the diffractive case, the resummation of initial state radiations in the TMD of the incoming proton strongly affects the azimuthal distribution of dijets. In experiments such as at the EIC, photoproduction contains both direct photon and resolved photon contributions. Here we focus on the direct photon contribution where the leading order partonic subprocesses are $\gamma g\to q\bar q$ and $\gamma q\to qg$. We take the example of the $\gamma g\to q\bar q$ channel. The other channel $\gamma q\to qg$ can be treated similarly. 

In $\gamma g\to q\bar{q}$, we will have contributions from  the soft gluon radiation associated with the incoming gluon. To lowest nontrivial order, the differential cross section  can be written as,
\begin{equation}
\frac{d^4\sigma}{d\Omega}=\sigma_0x_gf_g(x_g)\frac{1}{q_\perp^2}\left[\frac{\alpha_0^{\gamma g}}{\pi}+\frac{\alpha_2^{\gamma g}}{\pi}2\cos(2\phi)\right] \ ,\label{xsgp}
\end{equation}
where $\sigma_0$ represents the leading order cross section, $d\Omega=dy_1dy_2dP_\perp^2d^2q_\perp$ for the phase space, and $f_g(x_g)$ is the gluon distribution with $x_g=P_{\perp}\left(e^{ y_1}+e^{ y_2}\right)/\sqrt{S_{\gamma p}}$ is momentum fraction of the nucleon carried by the gluon. The coefficient $\alpha_{0,2}$ can be derived from the soft gluon radiation amplitude of~\cite{Mueller:2013wwa},
\begin{eqnarray}
\alpha_0^{\gamma g}&=&\frac{\alpha_s}{2\pi}\left[C_A\ln\frac{P_\perp^2}{q_\perp^2}+2C_F\ln\frac{a_0}{R^2} \right]\ ,\nonumber\\
\alpha_2^{\gamma g}&=&\frac{\alpha_s}{2\pi}\left[C_A\ln\frac{a_1}{R^2}-\frac{1}{N_c}\ln\frac{a_2}{R^2} \right]\ , \label{radiation}
\end{eqnarray}
where $a_1=1/e$ and $a_2$ is the same as  in the previous section. The emission from the incoming gluon is isotropic, and gives rise to the double logarithmic term $\frac{1}{q_\perp^2}\ln q_\perp^2$. 

We now perform the resummation of double logarithms in the standard TMD framework, 
\begin{eqnarray}
\frac{d^4\sigma}{d\Omega }&=&\sum_{ab}\sigma_0\int\frac{d^2\vec{b}_\perp}{(2\pi)^2}
e^{-i\vec{q}_\perp\cdot \vec{b}_\perp}\left[\widetilde{W}_{0}^{\gamma p}(|b_\perp|)\right.\nonumber\\
&&\left.~~-2\cos(2\phi_b)\widetilde{W}_{2}^{\gamma p}(|b_\perp|)\right]\ , \label{xc}
\end{eqnarray}
where the azimuthal symmetric term can be written as
\begin{eqnarray}
\widetilde{W}_0^{\gamma p}(b_\perp)=x_g\,f_g(x_g,\mu_b)
 e^{-S^{\gamma p}(P_\perp^2,b_\perp)}\  .\label{wb0}
\end{eqnarray}
We separate the Sudakov form factor $S(P_\perp,b_\perp)$ into perturbative and non-perturbative parts:
$S(P_\perp,b_\perp)=S_{\rm pert.}(P_\perp,b_\perp)+S_{\rm NP}(P_\perp,b_\perp)$ with the perturbative part at one-loop order is defined,
\begin{equation}
S_{\rm pert.}^{\gamma p}=\int^{P_\perp^2}_{\mu_b^2}\frac{d\mu^2}{\mu^2}\frac{\alpha_sC_A}{2\pi}
\left[\ln\frac{P_\perp^2}{\mu^2}-2\beta_0+\frac{2C_F}{C_A} \ln\frac{a_0}{R^2}\right]\ , \label{su}
\end{equation}
where $\beta_0=11/12-N_f/18$. For the non-perturbative part, we follow those for the TMD quark distributions in Refs.~\cite{Su:2014wpa,Prokudin:2015ysa} with an appropriate color factor,
\begin{equation}
    S_{\rm NP}^{\gamma p}=\frac{C_A}{C_F}\left[0.106\, b_\perp^2+0.42\ln\frac{P_\perp}{Q_0}\ln\frac{b_\perp}{b_*}\right] \ , \label{npgp}
\end{equation}
with $Q_0^2=2.4$~GeV$^2$.
For the $\cos(2\phi)$ term, we have 
\begin{eqnarray}
\widetilde{W}_2^{\gamma p}(b_\perp)=\alpha_2^{\gamma g}\widetilde{W}_0^{\gamma p}(b_\perp) \ .
\end{eqnarray}

\begin{figure}[tbp]
\begin{center}
\includegraphics[width=6.0cm]{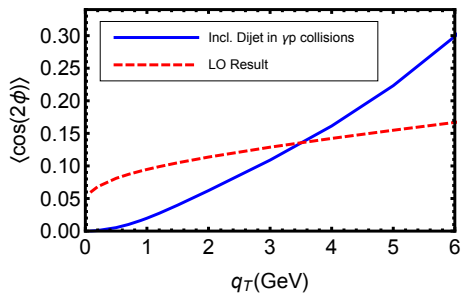}
\end{center}
\caption[*]{Anisotropy of inclusive dijet production in $\gamma p$ collisions at the future EIC for the typical kinematics: the leading jet $P_\perp=15~\rm GeV$ and both jets are at the same rapidity. Here we plot $\langle \cos(2\phi)\rangle$ as function and $q_\perp$, where $\phi$ is the azimuthal angle between $q_\perp$ and $P_\perp$. The leading order (LO) and full results are calculated from (\ref{radiation}) and (\ref{xc}), respectively. }
\label{fig:gpdijetv2qt}
\end{figure}

In Fig.~\ref{fig:gpdijetv2qt}, we show our numerical results for  $\langle \cos(2\phi)\rangle$ as a  function of $q_\perp$ in the leading order (from (\ref{radiation})) and resummed (from (\ref{xc})) cases.
We take the typical kinematics of  inclusive dijet production at the EIC with $\sqrt{S_{\gamma p}}=100\rm GeV$, $P_\perp\sim 15~\rm GeV$ and the two jets are at the same rapidity.  Though it is not manifest in the plot, the leading order (red) curve goes to zero as $q_\perp\to 0$, as it should. Compared to the results in the diffractive case, the impact of resummation is more pronounced. This is because the TMD resummation suppresses the region $q_\perp \sim 0$ and shifts the distribution of $q_\perp$ to larger values. This affects the angular dependent and independent parts of the cross section differently, and therefore their ratio $\langle \cos 2\phi\rangle$ has a stronger $q_\perp$-dependence. In contrast, in the diffractive case where the resummation is single-logarithmic, the effect largely cancels in the ratio.  

The $\cos(2\phi)$ asymmetry in inclusive dijet production in deep-inelastic scattering ($Q^2\neq 0)$ has been proposed to study the so-called linearly polarized gluon distribution in the nucleon~\cite{Boer:2010zf,Metz:2011wb,Dumitru:2015gaa,Boer:2016fqd}. As we have seen, the final state radiation from the dijet system  contributes  to exactly the same $\cos(2\phi)$ asymmetry already at leading order. Moreover, at higher orders the effect becomes  sensitive to the initial state radiation through the TMD resummation. It is important to carefully subtract these radiative contributions in order to unambiguously extract the linearly polarized gluon distribution from the experimental data. We will publish the relevant results in a separate paper.

{\it 4. Inclusive Dijet in $pp$ Collisions.}
We can also extend the above studies to inclusive dijet production in $pp$ collisions. Again we have to take into account the soft gluon radiation associated with incoming partons. In the following, we take the example of $gg\to gg$ channel, which is also the dominant contribution for dijet production in the typical kinematics at the LHC. The differential cross section and the resummation formula follow the similar expressions as Eqs.~(\ref{xsgp},\ref{xc},\ref{wb0}). For the soft gluon radiation contribution, we find the following results,
\begin{eqnarray}
\alpha_2^{g g}&=&\frac{\alpha_sC_A}{2\pi}\left[2\ln\frac{a_1}{R^2}
+\frac{\hat t^2+\hat u^2}{2(\hat s^2-\hat t\hat u)}\ln\frac{a_2}{a_1} \right]\ ,
\end{eqnarray}
and the associated $\alpha_0^{gg}$ has been derived in Ref.~\cite{Sun:2015doa}, where $\hat s$, $\hat t$ and $\hat u$ are usual Mandelstam variables in the partonic process.
For simplicity, we take the leading logarithmic approximation, for example, for the azimuthal symmetric one, 
\begin{equation}
\widetilde{W}_0^{gg}(b_\perp)=x_1\,f_g(x_1,\mu_b)
x_2\, f_g(x_2,\mu_b) e^{-S^{gg}(P_\perp^2,b_\perp)}\  ,\label{wb}
\end{equation}
where $f_{a,b}(x,\mu_b)$ are parton distributions for the incoming partons $a$ and $b$,  $x_{1,2}=P_{\perp}\left(e^{\pm y_1}+e^{\pm y_2}\right)/\sqrt{S}$ are momentum fractions of the incoming hadrons carried by the partons. 
The perturbative part is defined as one-loop order,
\begin{equation}
S_{\rm pert.}^{gg}=\int^{P_\perp^2}_{\mu_b^2}\frac{d\mu^2}{\mu^2}
\frac{\alpha_sC_A}{\pi}\left[\ln\left(\frac{P_\perp^2}{\mu^2}\right)-2\beta_0+\ln\frac{a_0}{R^2}\right]\ . \label{sugg}
\end{equation}
For the $\cos(2\phi)$ term, we have 
\begin{eqnarray}
\widetilde{W}_2^{gg}(b_\perp)=\alpha_2^{gg}\widetilde{W}_0^{gg}(b_\perp) \ .
\end{eqnarray}

\begin{figure}[tbp]
\begin{center}
\includegraphics[width=6.0cm]{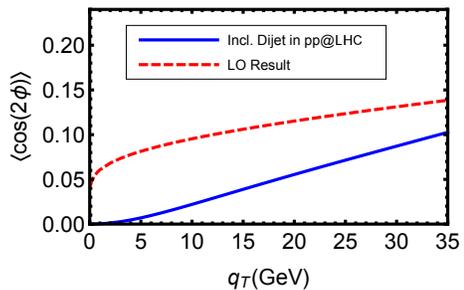}
\end{center}
\caption[*]{Anisotropy of inclusive dijet production from $gg\to gg$ channel in $pp$ collisions at the LHC at $\sqrt{S_{pp}}=7~\rm TeV$ for the typical kinematics: the leading jet $P_\perp=100\rm GeV$ and both jets are at mid-rapidity. Here we plot $\langle \cos(2\phi)\rangle$ as function and $q_\perp$, where $\phi$ is the azimuthal angle between $q_\perp$ and $P_\perp$. }
\label{fig:dijetv2qt}
\end{figure}

In Fig.~\ref{fig:dijetv2qt}, we show the $\langle \cos(2\phi)\rangle$ as function of $q_\perp$ for dijet production through $gg\to gg$ channel. In the numeric calculations, we take the typical kinematics at the LHC with $\sqrt{S_{pp}}=7~\rm TeV$, leading jet $P_\perp\sim 100\rm GeV$ and the two jets are both at mid-rapidity. 

In heavy ion collisions, the final state jet suffers multiple interactions with the hot QCD matter when it traverses through the medium. It also generates medium induced soft gluon radiations. Taking into account of the above two effects, we have the following modifications on the Fourier transform $\widetilde{W}$,
\begin{eqnarray}
\widetilde{W}_0&\Longrightarrow& \widetilde{W}_0^{med.}=\widetilde{W}_0+ \ Q_s^2 b_\perp^2/4,\\
\widetilde{W}_2&\Longrightarrow& \left(\alpha_2+\alpha_2^m\right)\widetilde{W}_0^{med.} \ ,
\end{eqnarray}
where $Q_s^2=\langle \hat q L\rangle $ represents the medium $p_T$-broadening effects with the medium transport coefficient $\hat q$ and jet traverse length $L$ and $\alpha_2^m$ for the medium induced soft gluon radiation contribution to the $\cos(2\phi)$ anisotropy. Because the medium induced radiation also has a collimation associated with the jet~\cite{Baier:1999ds}, this contribution will be similar to what was discussed in previous sections. Therefore, we may utilize the modification of the $\cos(2\phi)$ anisotropy as a measure of the medium induced soft gluon radiation in quark-gluon plasma in heavy-ion collisions.

{\it 5. Conclusions.}
In summary, we have studied the soft gluon radiation contribution to the dijet production in various hadronic collisions, focusing on the anisotropy of the total transverse momentum with respect to the leading jet. Our results have shown that these soft gluon radiation contributions lead to characteristic behaviors of $\langle \cos (2\phi)\rangle$ as functions of the total transverse momentum $q_\perp$. Experimental studies, as shown in a preliminary analysis of the diffractive dijet production in ultraperipheral $AA$ collisions at the LHC~\cite{CMS:2020ekd}, will provide a unique opportunity to investigate the final state radiation associated with the jet. Our finding has far-reaching consequences in extracting the linearly polarized gluon distribution and the elliptic Wigner distribution from the measurement of angular anisotropies in inclusive and exclusive dijet processes.  

A number of further developments shall follow. In this paper we focused on the $\cos(2\phi)$ anisotropy. Following the examples of Ref.~\cite{Catani:2017tuc}, other harmonics such as the $\cos(4\phi)$ anisotropy can be derived as well.  Second, we have made the narrow jet approximation in the derivations. The finite-$R$ corrections, together with the sub-eikonal corrections which might be relevant at large $q_\perp$, should be taken into account when we compare to the precise experimental data from the LHC and future EIC. Third, more rigorous derivation of the resummation formulas needs to be carried out along the line of Refs.~\cite{Catani:2014qha,Catani:2017tuc,Kang:2016mcy}. 

{\bf Acknowledgments:} We thank Daniel Tapia Takaki for discussions which partly motivated this work. We thank Stefano Catani and Massimiliano Grazzini for bringing our attention to their publications  \cite{Catani:2010pd,Catani:2014qha,Catani:2017tuc}, which are very relevant for the calculations in this paper. We also thank Felix Ringer for comments and discussions. Y.~H. thanks Yukawa institute, Kyoto university for hospitality.  This material is based upon work supported by the U.S. Department of Energy, Office of Science, Office of Nuclear Physics, under contract numbers DE-AC02-05CH11231 and DE- SC0012704. This work is also supported by the National Natural Science Foundations of China under Grant No. 11675093, No. 11575070 and Laboratory Directed Research and Development (LDRD) funds from Brookhaven Science Associates.

\end{document}